\documentclass[aps,showpacs,twocolumn,floatfix,bibnotes]{revtex4}

\usepackage{here}

\usepackage[dvips]{graphicx}

\newcommand\pictc[5]{\begin{figure}
                       \centerline{
                       \includegraphics[width=#1\columnwidth]{#3}}
                   \protect\caption{\protect\label{fig:#4} #5}
                    \end{figure}            }
\newcommand\pict[4][1.]{\pictc{#1}{!tb}{#2}{#3}{#4}}
\newcommand\rpict[1]{\ref{fig:#1}}

\newcommand\leqt[1]{\protect\label{eq:#1}}
\newcommand\reqtn[1]{\ref{eq:#1}}
\newcommand\reqt[1]{(\reqtn{#1})}

\newcounter{Fig}

\begin{document}
\begin{sloppy}

\title{Guided Modes in Negative Refractive Index Waveguides}

\author{Ilya V. Shadrivov}
\author{Andrey A. Sukhorukov}
\author{Yuri S. Kivshar}

\affiliation{Nonlinear Physics Group, Research School of Physical
Sciences and Engineering, Australian National University, Canberra
ACT 0200, Australia}
\homepage{http://www.rsphysse.anu.edu.au/nonlinear}

\begin{abstract}
We study linear guided waves propagating in a slab waveguide made
of a negative-refraction-index material, the so-called {\em
left-handed waveguide}. We reveal that the guided waves in
left-handed waveguides possess a number of peculiar properties, such as the absence of the fundamental modes, mode double degeneracy, and sign-varying
energy flux. In particular, we predict the existence of novel
types of guided waves with {\em a dipole-vortex structure} of the
Pointing vector.
\end{abstract}


\maketitle

Recent experimental demonstration of novel composite materials
with a negative index of refraction~\cite{Shelby:2001-77:SCI}
opens up a unique opportunity to design novel types of devices
where electromagnetic waves are governed in a non-conventional
way. The history of these materials begins with the paper by
Veselago \cite{Veselago:1967-517:UFN}, who studied the wave
propagation in a hypothetical material with simultaneously
negative dielectric permittivity $\epsilon$ and magnetic
permeability $\mu$. Such media are usually termed {\em
left-handed materials}, since the electric and magnetic fields
form a left set of vectors with the wave vector. Already back in
1968, Veselago predicted a number of remarkable properties of
waves in left-handed materials, such as negative refraction,
reversed Doppler and Vavilov-Cherenkov effects. However,
structures with both negative $\epsilon$ and $\mu$ have not been
known until recently, although materials with negative dielectric
permittivity are known (e.g. a metal below the plasma frequency).

The study of microstructured metallic materials for magnetic
resonance imaging \cite{Wiltshire:2001-849:SCI} has shown that
such structures can be described in terms of effective magnetic
permeability which becomes negative in the vicinity of a resonance
frequency. It was expected that mixing the composite materials
with negative magnetic permeability \cite{Wiltshire:2001-849:SCI}
with those possessing negative dielectric permittivity
\cite{Pendry:1996-4773:PRL} would allow to create a novel type of
{\em metamaterials} with a negative index of refraction. Indeed,
both numerical simulations \cite{Markos:2001-33401:PRB+} and
experimental results \cite{Shelby:2001-77:SCI,
Smith:2000-4184:PRL} confirmed that such left-handed (or
negative-index refraction) materials can be fabricated and
analyzed.

One of the first applications of the negative-refraction materials
was suggested by Pendry \cite{Pendry:2000-3966:PRL}, who
demonstrated that a slab of a lossless negative-refraction
material can provide a perfect image of a point source. Although
the perfect image is a result of an ideal theoretical model used
in Ref. \cite{Pendry:2000-3966:PRL}, the resolution limit was
shown to be independent on the wavelength of electromagnetic waves
(but can be determined by other factors such as loss, spatial
dispersion, etc.), and it can be better than the resolution of a
conventional lens \cite{Luo:2002-201104:PRB}.

The improved resolution of a slab of the negative-refraction
material can be explained by the excitation of surface waves at
both interfaces of the slab. Therefore, it is important to study
the properties of surface waves at the interfaces between the
negative-refraction and conventional materials. So far, the
frequency dispersion of surface waves at a single interface and in
a slab waveguide was calculated numerically only for particular
medium parameters~\cite{Ruppin:2000-61:PLA+}.

In this Letter, we study, for the first time to our knowledge, the
structure and basic properties of electromagnetic waves guided by
a left-handed waveguide. In order to emphasize the unusual and
somewhat exotic properties of such waves, we compare them with the
guided waves of conventional planar dielectric waveguides. We
reveal that the guided modes in left-handed waveguides differ
dramatically from conventional guided waves, and they possess a
number of unusual properties, including the absence of the
fundamental modes, double degeneracy of the modes, the
sign-varying energy flux, etc. In particular, we predict the
existence of novel types of guided waves with a dipole-vortex
structure of the energy flux and the corresponding Pointing
vector.

\pict{fig01.eps}{dispersion}{ Comparison between the conventional
(a)~and left-handed (b)~guided modes of a slab waveguide. The
dashed and solid curves correspond to the left- and right-hand
sides of the dispersion relations in Eqs.~\reqt{dispersion}
and~\reqt{dispersionI}, respectively. Intersections of these
curves indicate the existence of guided modes. Three dashed lines
in each plot correspond to waveguides with different parameters
$\rho_1> \rho_2 > \rho_3$, but the fixed ratio $\mu_2/\mu_1$. Inserts show
the waveguide geometry, and the transverse profiles of the guided
modes.}

We consider a symmetric slab waveguide in a conventional planar
geometry [see, e.g., the top left insert in
Fig.~\rpict{dispersion}(a)]. In the general case, a slab of the
thickness $2 L$ is made of a material with dielectric permittivity
$\epsilon_2$ and magnetic permeability $\mu_2$, which both can be
negative or positive. We assume that the surrounding medium is
right-handed, and is therefore characterized by both positive
$\epsilon_1$ and $\mu_1$. It is well known that a slab waveguide
made of a conventional (right-handed) dielectric material with
$\epsilon_2 > 0$ and $\mu_2 > 0$ creates a non-leaky waveguide for
electromagnetic waves, provided the refractive index of a slab is higher than that of the surrounding dielectric medium (cladding), i.e. $\epsilon_2 \mu_2 >
\epsilon_1 \mu_1$. However, in the following we demonstrate that
this simple criterion can not be applied to the waveguides made of
a left-handed material.

\pict{fig02.eps}{frequency}{ Frequency dispersion curves for the
three lowest-order guided modes of the left-handed slab waveguide
($L=2$ cm). Insets show the characteristic mode profiles.}

To be specific, below we describe the properties of the
TE guided modes in which the electric field $\vec{E}$ is polarized along
the $y$ axis. A similar analysis can be carried out for the TM
modes, and these results will be presented elsewhere. From the
Maxwell's equations, it follows that stationary TE modes can be
described by the following equation for the scalar electric field
$E = E_y$,
\begin{equation} \leqt{Helm}
   \left[ \frac{\partial^2}{\partial z^2}
          + \frac{\partial^2}{\partial x^2}
          + \frac{\omega^2}{c^2} \epsilon(x) \mu(x)
          - \frac{1}{\mu(x)} \frac{\partial \mu}{\partial x}
            \frac{\partial}{\partial x} \right] E
   = 0,
\end{equation}
where $\omega$ is the angular frequency of the monochromatic
waves.

The guided modes can be found as special stationary solutions of
Eq.~\reqt{Helm} having the following form,
\begin{equation} \leqt{mode}
   E(x,z) = E_0(x) e^{i h z} ,
\end{equation}
where real $h$ is the wave propagation constant and $E_0(x)$ is
the spatially localized transverse profile of the mode. After substituting
Eq.~\reqt{mode} into Eq.~\reqt{Helm}, we obtain an eigenvalue
problem that possesses spatially localized solutions for
\begin{equation} \leqt{kappa1}
   \kappa_1^2 = h^2 - \frac{\omega^2}{c^2}\epsilon_1\mu_1 > 0 ,
\end{equation}
because only in this case the mode amplitude decays away from the
waveguide, $E_0(x) \sim \exp( -|x| \kappa_1)$.

We solve the eigenvalue problem in each of the homogeneous layers,
and then employ the corresponding boundary conditions following
from Eq.~\reqt{Helm}. As a result, we obtain the dispersion
relations which define a set of allowed eigenvalues $h$,
\begin{equation} \leqt{dispersion}
   \left(\kappa_1 L \right) = \pm \frac{\mu_1}{\mu_2}(k_2 L) \,
                                  \tan^{\pm 1}(k_2 L),
\end{equation}
where the signs $(+)$ and $(-)$ correspond to the symmetric and
antisymmetric guided modes, respectively, and $k_2 =
[(\omega^2/c^2) \epsilon_2 \mu_2 - h^2]^{1/2}$. When $k_2$ is
real, the corresponding modes can be identified as ``fast waves'',
since their phase velocity $\omega/h$ is larger than the phase
velocity in an homogeneous medium with $\epsilon_2$ and $\mu_2$.

The parameter $k_2$ becomes purely imaginary for ``slow waves'',
when the propagation constant $h$ exceeds a critical value. Then,
it is convenient to present Eq.~\reqt{dispersion} in an equivalent
form using $\kappa_2 = i k_2$,
\begin{equation} \leqt{dispersionI}
   \left(\kappa_1 L \right) = - \frac{\mu_1}{\mu_2}(\kappa_2 L) \,
                                  \tanh^{\pm 1}(\kappa_2 L) .
\end{equation}

Following a standard textbook analysis (see, e.g.
Ref.~\cite{Vinogradova:1990:TheoryWaves}), we consider the
parameter plane $\left(k_2 L, \kappa_1 L \right)$, and also extend
it by including the imaginary values of $k_2$ using the auxiliary
parameter plane $\left(\kappa_2 L, \kappa_1 L \right)$. In
Figs.~\rpict{dispersion}(a,b), we plot the dependencies described
by the left-hand (dashed) and right-hand (solid) sides of
Eqs.~\reqt{dispersion} and~\reqt{dispersionI}, using a relation
between the parameters,
\begin{equation} \leqt{dispCircle}
   \left( \kappa_1 L \right)^2 + \left( k_2 L \right)^2 
   = L^2 \left( \omega^2/c^2\right) 
     \left( \epsilon_2 \mu_2 - \epsilon_1 \mu_1 \right) 
   \equiv \rho .
\end{equation}
In Figs.~\rpict{dispersion}(a,b), we draw three dashed lines
corresponding to different slab waveguides, having the same ratio
$\mu_1/\mu_2$. The intersections of a dashed line with solid
curves indicate the existence of various guided modes. We present
results for a conventional (right-handed) waveguide in
Fig.~\rpict{dispersion}(a), in order to compare them directly with the
corresponding dependencies for a left-handed slab waveguide in
Fig.~\rpict{dispersion}(b).

First of all, the analysis of Eqs.~\reqt{dispersion}
and~\reqt{dispersionI} confirms the well-known textbook results that a right-handed slab waveguide can only support ``fast'' guided modes, which exist when the waveguide core has a higher refractive index than its cladding, i.e. for $\epsilon_2 \, \mu_2 > \epsilon_1 \, \mu_1$. In
this case, there always exists {\em a fundamental guided mode}, which
profile does not contain zeros. The conventional waveguide can
also support higher-order modes, their number is limited by the
value $2 \rho^{1/2} / \pi$. These various regimes are
illustrated in Fig.~\rpict{dispersion}(a) with different dashed
lines.

However, we find that the properties of the left-handed slab
waveguides are highly nontrivial. First, such waveguides can
support ``slow'' modes, and they are either symmetric (node-less)
or antisymmetric (one zero). Such solutions represent in-phase or
out-of-phase bound states of surface modes, localized at the two
interfaces between right and left media. In the conventional case
of both positive $\epsilon$ and $\mu$, such surface waves do not
exist, however they appear when the magnetic permeability changes
its sign (for the TE polarization). Thus, the guided modes can be
supported by {\em both low-index and high-index} left-handed slab
waveguides.

Second, the conventional hierarchy of ``fast'' modes is removed.
Specifically, (i)~the fundamental node-less mode does not exist at
all, (ii)~the first-order mode exists only in a particular range
of the parameter values $\rho$, and it always disappears in
wide waveguides, when $\rho$ exceeds a critical value, and
(iii) two modes having the same number of nodes can
co-exist in the same waveguide. We illustrate some of these
nontrivial features in Fig.~\rpict{dispersion}(b).

\pict{fig03.eps}{widthTuning}{
Surface waves in a slab waveguide for the
case $\epsilon_2\mu_2 < \epsilon_1\mu_1$ and $\mu_2<\mu_1$. Shown are
(a)~the propagation constant and (b)~the energy flux vs. the slab
thickness parameter $L$. Solid and dashed lines correspond to strongly and weakly localized modes, respectively. Insets show the characteristic mode profiles.}

Frequency dispersion of the guided waves in the left-handed
waveguides should be studied by taking into account the dispersion
of both $\epsilon_2$ and $\mu_2$, since this is an essential
property of such materials~\cite{Veselago:1967-517:UFN}. We follow
Ref.~\cite{Smith:2000-4184:PRL} and consider the following
frequency dependencies of the effective left-handed medium
characteristics,
\begin{equation} \leqt{freqDisp}
   \epsilon_2(\omega) = 1 - \frac{\omega_p^2}{\omega^2} , \;\;\;\;\;
   \mu_2(\omega) = 1 - \frac{F\omega^2}{\omega^2-\omega_0^2} ,
\end{equation}
where the parameters correspond to the experimental data of
Ref.~\cite{Smith:2000-4184:PRL}: $\omega_p = 10$~GHz, $\omega_0 =
4$~GHz, and $F = 0.56$. The region of simultaneously negative
permittivity and permeability in this case ranges from 4 GHz to 6
GHz. Dispersion curves for the first three guided modes in a slab waveguide with the thickness parameter $L = 2$~cm are shown in
Fig.~\rpict{frequency}, where dashed curves correspond to ``fast''
modes, and solid -- to ``slow'' modes. We find that the fundamental
``slow'' mode exists only at higher frequencies, whereas the
second-order fast mode appears at lower frequencies. Both modes
can have either positive or negative group-velocity dispersion in
different parameter regions. Properties of the first-order antisymmetric mode are different. The type of this mode seamlessly changes from ``fast'' to ``slow'' as the frequency grows. This transition
occurs when the condition $k_2=0$ is satisfied, which is a boundary
separating the two types of modes, as shown in Fig.~\rpict{frequency}
by a dotted line. The high ``fast'' modes exists at the frequencies close to the resonance at $\omega=4GHz$.

In the left-handed materials, the electromagnetic waves are
backward, since the energy flux and wave vector have opposite
directions~\cite{Veselago:1967-517:UFN}, whereas these vectors are
parallel in conventional (right-handed) homogeneous materials. The
energy flux is characterized by the Pointing vector averaged over
the period $T=2\pi / \omega$ and defined as ${\bf S} = (c / 8 \pi)\,{\rm
Re}\left[ {\bf E}\times {\bf H}^\ast \right]$. A monochromatic
guided mode has, by definition, a stationary transverse profile,
and the averaged energy flux is directed along the waveguide only.
It follows from the Maxwell's equations and Eq.~\reqt{mode} that
the $z$-component of the energy flux is,
\begin{equation} \leqt{flux}
   S_z = \frac{c^2 h}{8 \pi \omega \mu(x)} E_0^2 .
\end{equation}
The total power flux through the waveguide core and cladding can
be found as $P_2 = \int_{-L}^{L} S_z\,dx$ and $P_1 = 2\;
\int_{L}^{+\infty} S_z\,dx$, respectively. We find that the
energy flux distribution for the waves guided along the
left-handed slab is rather unusual. Indeed, the energy flux inside
the slab (with $\mu<0$) is opposite to that in the surrounding
medium (with $\mu>0$). This occurs because the normalized wave
vector component along the waveguide ($h$) is fixed in a guided
mode according to Eq.~\reqt{mode}. An important information about
the guided modes can be extracted from the study of the normalized
energy flux $P=(P_1+P_2)/(|P_1|+|P_2|)$. This parameter is
bounded, $|P| < 1$, $P \rightarrow 1$ when the mode is weakly
localized ($|P_1|\gg|P_2|$), whereas $P < 0$ for modes which are
highly confined inside the left-handed slab.

We have performed a detailed analysis of the slow guided modes and
identified four distinct cases.

(i) $\epsilon_2\mu_2 > \epsilon_1\mu_1$, $\mu_2>\mu_1$. Only odd
mode exists below the threshold,  $\rho <  \mu_1^2 / \mu_2^2$. The
corresponding critical value of the slab thickness $L$ below which
the odd mode exists is found as
\begin{equation} \leqt{Lcrit}
   L_{\rm cr} = \frac{c}{\omega} 
                \frac{\mu_1}{\mu_2 \sqrt{\epsilon_2\mu_2-\epsilon_1\mu_1}}.
\end{equation}
The energy flux $P$ is positive for all values of $L$. The modes
are forward propagating, i.e. the total energy flux along the waveguide
is co-directed with the wavevector.

\pict{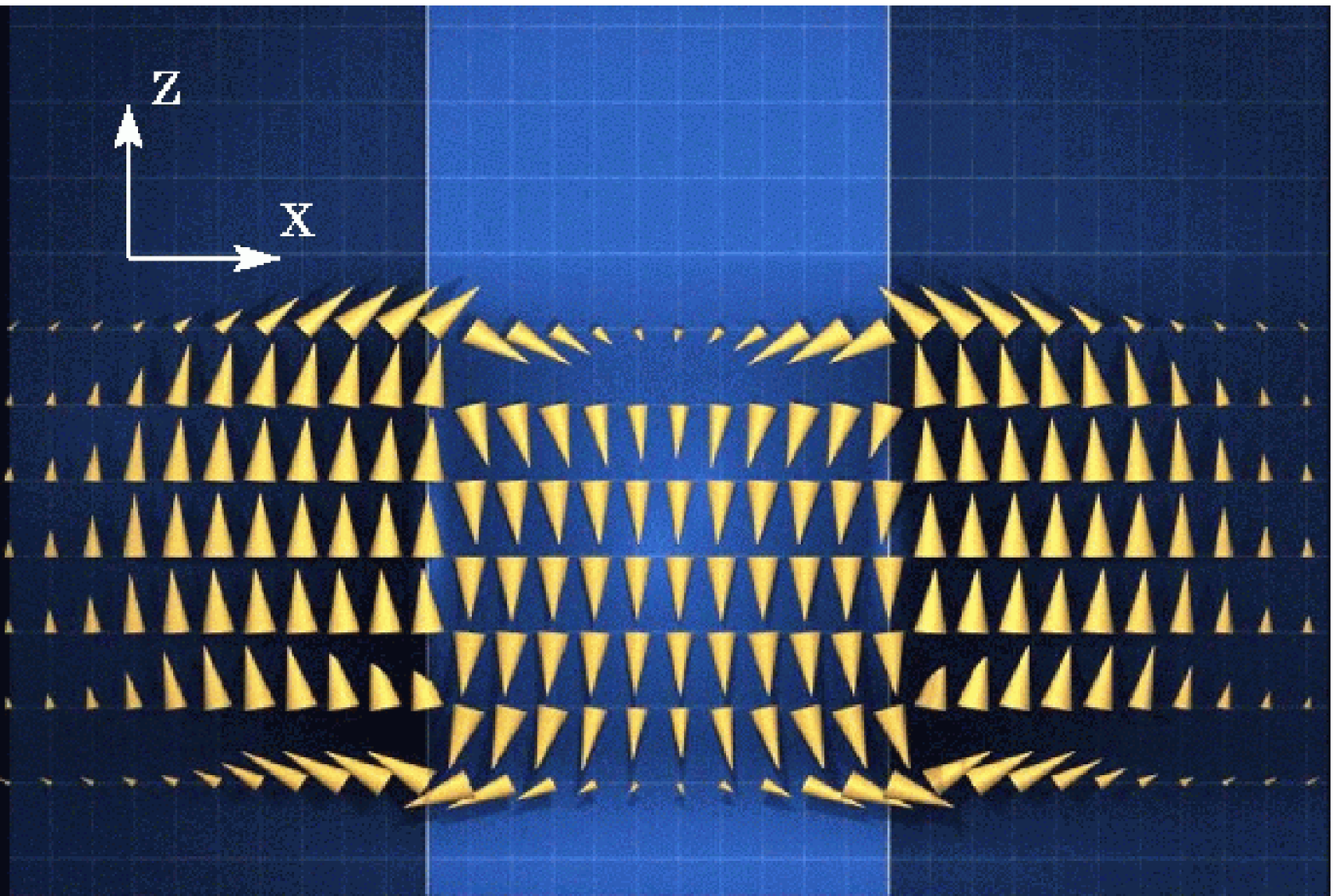}{vortex}{ Structure of the Pointing vector field in a
localized surface wave propagating along a left-handed slab.}

(ii) $\epsilon_2\mu_2 > \epsilon_1\mu_1$, $\mu_2 < \mu_1$. Even
mode exists at all values of $\rho$, however odd modes can appear only when a threshold parameter value is exceeded, $\rho > \mu_1^2/\mu_2^2$. Accordingly, the critical value~\reqt{Lcrit} determines the
lower boundary of the existence region for odd modes. The total
energy flux is negative for all $L$, and the modes are backward.
The energy is mostly localized inside the slab.

(iii) $\epsilon_2\mu_2<\epsilon_1\mu_1$, $\mu_2>\mu_1$. Both odd
and even modes exist at all values of $\rho$ and $L$, and the
modes are forward.

(iv) $\epsilon_2\mu_2 < \epsilon_1\mu_1$, $\mu_2<\mu_1$. Only even
modes exist below the threshold value of $\rho$ which can be found
numerically using Eq.~\reqt{dispersionI}. Characteristic
dependences of the wavenumber and normalized power on the slab
width is shown in Figs.~\rpict{widthTuning}(a,b). At any slab
thickness below a critical value, two modes always co-exist. One
of the modes is forward and weakly localized, but the other
one is backward and more confined. When the slab width
approaches the critical value, the branches corresponding to
different modes merge, and the energy flux vanishes. In this
special case, the energy fluxes inside and outside the slab
exactly compensate each other.

Since the energy fluxes are oppositely directed inside the guided
modes, it might initially seem that such waves can only be
sustained by two continuously operating emitters positioned at the
opposite ends of the waveguide. Therefore, it is of the
fundamental importance to understand whether wave packets of
finite temporal and spatial extension can exist in left-handed
waveguides. We calculate the Pointing vector averaged over the
period of the carrier frequency, and present the characteristic
structure of the energy flow in Fig.~\rpict{vortex}. Due to the
unique double-vortex structure of the energy flow, most of the
energy remains inside the wavepacket, and it does not disintegrate. The group velocity is proportional to the total energy flux $P$, and it can therefore be made very small or even zero by a proper choice of the waveguide parameters as demonstrated above. On the other hand, the group-velocity dispersion, which determines the rate of pulse broadening, can also be controlled. This flexibility seems very promising for
potential applications.

Finally, we note that recent numerical simulations demonstrated
that the phenomenon of the negative refraction, similar to that
found for the left-handed metamaterials, can be observed in {\em
photonic crystals}~\cite{Luo:2002-201104:PRB,Kosaka:1998-10096:PRB+}.
Although in this case the wavelength is of the same order as the
period of the dielectric structure (and, therefore,
a simple analysis in terms of the effective medium approximation
is not strictly justified), we expect that similar mechanisms of
wave localization will remain generally valid.

In conclusion, we have described, for the first time to our
knowledge, linear guided waves in left-handed slab waveguides. We
have demonstrated a number of exotic properties of such waves,
including the absence of fundamental modes and the sign-varying
energy flux, and we have predicted the existence of the
fundamentally novel classes of guided waves with a vortex-type
internal structure.

We thank C.~T. Chan and C.~M. Soukoulis for
useful discussions, and to D.~E. Edmundson for assistance with Fig.~\rpict{vortex}. The work was partially supported by the Australian Research Council.

\label{Markos:2001-33401:PRB+, Smith:2002-195104:PRB}
\label{Kosaka:1998-10096:PRB+, Natomi:2002-133:OQE}
\label{Ruppin:2000-61:PLA+, Ruppin:2001-1811:JPCM, Haldane:cond-mat/0206420:ARXIV}

\end{sloppy}

\begin{thebibliography}{99}

\bibitem{Shelby:2001-77:SCI}
R.A. Shelby, D.R. Smith, and S. Shultz, Science
{\bf 292}, 77 (2001).

\bibitem{Veselago:1967-517:UFN}
V.G. Veselago, Usp. Fiz. Nauk {\bf 92}, 517 (1967) [Sov. Phys.
Usp. {\bf 10}, 509 (1968)].

\bibitem{Wiltshire:2001-849:SCI}
M.C.K. Wiltshire, J.B. Pendry, I.R. Young, D.J. Larkman, D.J.
Gilderdale, and J.V. Hajnal, Science {\bf 291}, 849 (2001).

\bibitem{Pendry:1996-4773:PRL}
J.B. Pendry, A.J. Holden, W.J. Stewart, and I. Youngs, Phys. Rev.
Lett. {\bf 76}, 4773 (1996).

\bibitem{Markos:2001-33401:PRB+}
P. Marko\v{s} and C.M. Soukoulis, Phys. Rev. B {\bf 65}, 033401
(2001);  D.R. Smith, S. Schultz, P. Marko\v{s}, and C.M.
Soukoulis, Phys. Rev. B {\bf 65}, 195104 (2002).

\bibitem{Smith:2000-4184:PRL}
D.R. Smith, W. Padilla, D.C. Vier, S.C. Nemat-Nasser, and S.
Shultz, Phys. Rev. Lett. {\bf 84}, 4184 (2000).

\bibitem{Pendry:2000-3966:PRL}
J.B. Pendry, Phys. Rev. Lett. {\bf 85}, 3966 (2000).

\bibitem{Luo:2002-201104:PRB}
C. Luo, S.G. Johnson, J.D. Joannopoulos, and J.B. Pendry, Phys.
Rev. B {\bf 65}, 201104 (2002).

\bibitem{Ruppin:2000-61:PLA+}
R. Ruppin, Phys. Lett. A {\bf 277}, 61 (2000);
J. Phys. Condens. Matter {\bf 13}, 1811 (2001);
F.D.M. Haldane, arXiv:cond-mat:0206420 (2002).

\bibitem{Vinogradova:1990:TheoryWaves}
M.~B. Vinogradova, O.~V. Rudenko, and A.~P. Sukhorukov, {\em The
Theory of Waves} (Nauka, Moscow, 1990) (in Russian).

\bibitem{Kosaka:1998-10096:PRB+}
H. Kosaka, T. Kawashima, A. Tomita, M. Notomi, T. Tamamura, T.
Sato, and S. Kawakami, Phys. Rev. B {\bf 58}, 10096 (1998); M.
Natomi, Opt. and Quant. Electron. {\bf 34}, 133 (2002).


\end{thebibliography}
\end{document}